\documentclass{article}

\usepackage{arxiv}

\usepackage[utf8]{inputenc}
\usepackage{amsmath,amssymb,bbm}
\usepackage{relsize}
\usepackage{graphicx}
\usepackage{arydshln}
\usepackage{booktabs}
\usepackage{multirow}
\usepackage{natbib}
\usepackage{float}
\usepackage[flushleft]{threeparttable}
\usepackage{setspace}

\newcommand{\te}[1]{\text{#1}}
\newcommand{\ti}[1]{\textit{#1}}

\newcommand{\E}[2]{\mathbbm{E}_{#1}\left[{#2}\right]}

\newcommand\numberthis{\addtocounter{equation}{1}\tag{\theequation}}

\title{Maximum a Posteriori Estimation of Dynamic Factor Models with Incomplete Data}

\title{Maximum a Posteriori Estimation of Dynamic Factor Models with Incomplete Data}

\author{
  Erik Sp{\aa}nberg \\
  Department of Statistics, Stockholm University \\ SE-106 91, Stockholm \\ E-mail: erik.spanberg@stat.su.se
}
\bibpunct[, ]{(}{)}{;}{a}{,}{,}

\begin{document}

\maketitle
\onehalfspacing
\begin{abstract}
In this paper, we present a method of maximum a posteriori estimation of parameters in dynamic factor models with incomplete data. We extend maximum likelihood expectation maximization iterations by \cite{BanburaetAl2014} to penalized counterparts by applying parameter shrinkage in a Minnesota prior style fashion, also considering factors loading onto variables dynamically.  The missing data is refined by preemptive integration over absent observations. Additionally, a heuristic and adapting shrinkage scheme is considered. The algorithm is applicable to any arbitrary pattern of missing data, including different publication dates, sample lengths and frequencies. The method is evaluated in a Monte Carlo study, generally performing favorably, and at least comparably, to maximum likelihood.
\end{abstract}

% keywords can be removed
\keywords{Dynamic factor model \and Maximum a posteriori estimation \and Missing data \and Expectation Maximization}

\section{Introduction}
This paper introduces an algorithm for maximum a posteriori (MAP) estimation of parameters in dynamic factor models with incomplete data, suitable for any arbitrary case of missing data pattern.

A factor model's fundamental task is to isolate the main co-movement of a (possibly large) number of at least partially observable variables into an unobservable, however estimable,  \textit{common component}. The remaining individual variable movements are called \textit{idiosyncratic components}. By assuming that the common component is driven by just a few factors, a large system of variables can be summarized by a smaller dimension, gaining parsimony in structural analysis and forecast applications. 

\textit{Dynamic} factor models (DFMs) have gained substantial traction in econometric practice ever since some early contributors \citep[in particular][]{SargentSims1977, Geweke1977, GewekeSingleton1981, EngleWatson1981}. Such applications include construction of summarizing economic indicators \citep[e.g.][]{StockWatson1989, FraleEtAl2011}, analyzing business cycles \citep[e.g.][]{ForniReichlin1998, HuberEtAl2020}, monetary policy analysis \citep[e.g.][]{BernankeBoivin2003, BoivinEtAl2009}, yield curve estimation \citep[e.g.][]{DieboldEtAl2006, LengwilerCarlos2010} and macro-economic forecasting \citep[e.g.][]{StockWatson2002a,StockWatson2002b}. 

DFMs have also elicited a particular interest among macroeconometricians dealing with missing data. This is in part due to the so called "ragged edge" issue. Ragged edge refer to a particular unbalanced panel of cross sections, with missing data in some of the observable sample endpoints. This a common pattern of missing data in practical forecasting settings,  as macro-economic indicators often have different publication delays. The DFM has been deemed an efficient tool in handling patterns of this kind, making it a central piece in macro-economic 'nowcasting' literature \citep[e.g.][]{GiannoneEtAl2008, BanburaRunstler2011, BanburaEtAl2011, DozEtAl2011, SolbergerSpanberg2020}. 

\cite{DempsterEtAl1977} introduced the Expectation Maximization (EM) algorithm to deal with missing data in Maximum likelihood (ML) estimation. Further, \cite{ShumwayStoffer1982} show how the EM algorithm also can be applied in case of latent states in a state space representation using the Kalman filter and smoother. As it happens, DFMs can be written in state space representations where factors are latent states \citep[see][for an early example]{EngleWatson1981}. \cite{WatsonEngle1983} utilize this fact to apply the EM-algorithm to dynamic factor model estimation, but only in case of complete data. \cite{BanburaetAl2014} extend their result to the incomplete data case, allowing for any arbitrary missing data pattern. Such patterns include, but are not limited to, ragged edge, variables starting at different dates, mixed frequencies and data imputation errors. Their approach can therefore be useful in a range of different situations, including real time economic forecasting in emerging countries or local economies, where some data might be scarce or in any way incomplete. 

In this paper we extend the EM-iterations of \cite{BanburaetAl2014} to a Bayesian counterpart, while dealing with additional flexibility in model dynamics. Following the definitions of \cite{BaiNg2007}, the dynamic factor model assumes the common component to be dynamic in two ways: first, the factors follow a dynamic process; second, the factors load onto the observable series dynamically. Several estimation methods dealing with missing data,  including \cite{BanburaetAl2014}, only handle dynamics in the first sense \citep[see also][]{ GiannoneEtAl2004, DozEtAl2011, JungbackerEtAl2011}. We derive an estimation method which allows for dynamics in the double sense. As such, we utilize potentially diverse lag or lead cross-correlations between series, possibly increasing predictive capabilities, at the cost of introducing more parameters and consequently greater risk of overfitting. To deal with this risk, we introduce parameter shrinkage by prior distributions similar to what is commonly used in Bayesian Vector Autogression (VAR) models \citep[see][]{DoanetAl1984, Litterman1986, KadiyalaKarlsson1997, BanburaetAl2010, Karlsson2013}.

To summarize, we propose a method which can deal with many different cross-section sizes and cross-correlations with missing data. We draw insights from EM-algorithm literature to deal with missing data, and combine insights from factor model and BVAR literature to deal with large cross-sections. 

The rest of the paper is outlined as follows. Section 2 describes the model framework with Minnesota style priors and corresponding EM-iterations for MAP estimation. Additionally, we show a heuristic, but quick, way of choosing hyperparameters. Section 3 evaluates the method's estimation capabilities in a Monte Carlo exercise. Section 4 concludes. 

\section{Model framework}
Let $\ti{y}_t = [\ti{y}_{1,t} \; \ti{y}_{2,t} \; ... \; \ti{y}_{n,t}]'$ be a vector of $n$ time series at time $t$, from a multivariate weakly stationary process $\{\ti{y}_t, t \in \mathbb{Z}\}$  with mean zero, $\E{}{\ti{y}_{i,t}} = 0 \; (i = 1, ..., n)$, and finite second order moments $\E{}{\ti{y}_{i,t}\ti{y}_{i,t-g}} < \infty \; (g \in \mathbb{Z})$. We assume the specification:
\begin{align}
        \ti{y}_t &= \Lambda_0 f_t + \Lambda_1 f_{t-1} + ... + \Lambda_p f_{t-p} + \epsilon_t, \quad &\epsilon_t \overset{\mathit{iid}}{\sim} \mathcal{N}\left(0, \Psi^{-1}\right) \label{ObserveForm1}
\end{align} for $t=1, ..., T$, where $f_t = [f_{1,t} ... f_{r,t}]$ is a vector of $r$ dynamic common factors and $\epsilon_t$ is a $n \times 1$ vector of idiosyncratic components. Parameters are collected in $n \times r$ matrices $\Lambda_0, ..., \Lambda_p$ of factor loadings and $n$ positive idiosyncratic precisions (inverse variances) placed in a  diagonal matrix $\Psi = \text{diag}(\psi_1, ..., \psi_n)$. As $\Psi$ is assumed diagonal, this represents an exact factor model. Observe that the factors load onto the variables dynamically. Moreover, the factors belong to a vector autoregressive process according to
\begin{align}
    f_t &= \Phi_1 f_{t-1} + ... + \Phi_q f_{t-q} + \xi_t, \quad &\xi_t \overset{\mathit{iid}}{\sim} \mathcal{N}\left(0, \Omega^{-1}\right) \label{Factorform1}
\end{align} where $\xi_t$ is a $r \times 1$ vector of residuals with precisions $\Omega = \text{diag}(\omega_1, ..., \omega_r)$ and $\Phi_1, ..., \Phi_q$ are $r \times r$ matrices of parameters.  Additionally, we assume the processes $\{\epsilon_{t}\}_{t=1}^\infty$ and $\{\xi_t\}_{t=1}^\infty$ to be independent. The factors can be written in a Wold representation, where they originate from \textit{primitive common shocks} \citep[see][]{ForniEtAl2000}.

\eqref{ObserveForm1}-\eqref{Factorform1}
are written more condensed as
% Model formulation, stacked
\begin{align}
    \ti{y}_t &= \Lambda F^{\Lambda}_t + \epsilon_t,  & &\epsilon_t \sim \mathcal{N}\left(0, \Psi^{-1}\right) \label{ObserveForm2} \\
    f_t &= \Phi F^{\Phi}_{t-1} +  \xi_t, & &\xi_t \sim \mathcal{N}\left(0, \Omega^{-1}\right) \label{FactorForm2}
\end{align}

with stacked matrices and vectors

% Objects in model formulation
\begin{align}
    \Lambda &= \begin{bmatrix} \Lambda_0 & \Lambda_1 & ... & \Lambda_p \end{bmatrix} \\
    \Phi &= \begin{bmatrix} \Phi_1 & \Phi_2 & ... & \Phi_q \end{bmatrix}\\
    F_t^\Lambda &= \begin{bmatrix} f_t' & f_{t-1}' & ... & f_{t-p}' \end{bmatrix}' \\
    F_{t-1}^\Phi &= \begin{bmatrix} f_{t-1}' & f_{t-2}' & ... & f_{t-q}' \end{bmatrix}' \label{Fphi}
\end{align} The common component is given by $\mathcal{X}_t = \Lambda F^{\Lambda}_t$, defined by a lag-polynomial operation on the factors.

\subsection{Minnesota style priors}

We adopt a Minnesota style prior in the spirit of \cite{Litterman1986} \citep[see also][]{KadiyalaKarlsson1997,SimsZha1998}. The original suggestion of \cite{Litterman1986} was to a-priori center all equations around random walks with drift, arguing that macroeconomic time series often act like drifting first-order integrated stochastic processes. For stationary variables with mean zero that would correspond to center all equations to white noise. This belief is practically enforced by applying shrinkage to the coefficients.

Moreover, they incorporate prior beliefs that more recent lags are more important than distant lags, introducing so called \textit{lag-decay}. Technically, this is done by increasing the shrinkage of the coefficients with each lag they correspond to. 
We adopt similar a procedure for dynamic factor models,  centering all series and factors to independent white noise by setting prior moments:
\begin{align}
\mathbbm{E}\left[(\Lambda_\ell)_{i,z}\right] &= 0, \quad & \mathbbm{E}\left[(\Phi_\ell)_{i,z}\right] &= 0, \\
      \mathbbm{V}\left[(\Lambda_\ell)_{i,z}\right] &= \begin{cases} \mathlarger{\frac{1}{\eta_{\lambda_i}(\ell + 1)^{d_\Lambda}}}, &\text{ when } i = z \\ \\ 0,  &\text{ when } i \neq z\end{cases},
     \quad & \mathbbm{V}\left[(\Phi_\ell)_{j,z}\right] &= \begin{cases} \mathlarger{\frac{1}{\eta_\Phi \ell^2}}, &\text{ when } j = z \\ \\ 0,  &\text{ when } j \neq z \end{cases},
\end{align}
where $\mathbbm{V}[\cdot]$ denotes the variance.

Every individual parameter in $\Lambda$ and $\Phi$ are assumed to be a-priori independent and Gaussian distributed. Let us denote $\lambda_i'$ as the $i$th row of $\Lambda$ and $\phi_j'$ as the $j$th row of $\Phi$. Then $\eta_{\lambda_i}$ and $\eta_{\phi_j}$ decide their overall shrinkage respectively, with greater values meaning higher shrinkage. The factors $1/(\ell+1)^{d_\Lambda}$ and $1/\ell^{d_\Phi}$ define the rate of which shrinkage increases by lag length, decided by the lag-decay parameters $d_\Lambda$ and $d_\Phi$. The prior can be described as
% Priors
\begin{align}
    \lambda_i &\sim \mathcal{N}\left(0,V_i\right), \quad i=1, ...,n \\
    \phi_j &\sim \mathcal{N}\left(0,W_j\right), \quad j=1, ..., r, 
\end{align} where 
\begin{align}
    V_i &= \frac{1}{\eta_{\lambda_i}} J_\Lambda^{-1}, &\quad  J_\Lambda &= \begin{bmatrix} 1 & 2^2 & ... & (p+1)^2 \end{bmatrix} \otimes I_r, \\ 
    W_j &= \frac{1}{\eta_{\phi_j}} J_\Phi^{-1}, &\quad  J_\Phi &=  \begin{bmatrix} 1 & 2^2 & ... & q^2 \end{bmatrix} \otimes I_r.
\end{align}

Further we apply a diffuse prior for precisions given by
\begin{align*}
        \psi_i &\sim \frac{1}{\sqrt{\psi_i}}, \quad i = 1, ..., n \numberthis \label{priorvarpsi} \\
        \omega_j &\sim \frac{1}{\sqrt{\omega_j}}, \quad j = 1, ..., r \numberthis \label{priorvaromega}
\end{align*}  These are not a proper prior distributions, but can serve as approximations of suitable diffuse proper priors. They are however informative, in the sense that it will influence the EM-iterations relative to the ML case, which will be shown. 

Furthermore we apply a Gaussian prior on initial factors to initialize the Kalman filter by:
\begin{align}
    \widetilde{F}_0 \sim \mathcal{N}\left(0, \widetilde{\Sigma}_0\right)
\end{align}
where $\widetilde{F}_0 = \begin{bmatrix} f_0' & f_{-1}' & ... & f_{-s}' \end{bmatrix}'$, $s = \max(p, q-1)$ and $\widetilde{\Sigma}_0$ is a diagonal matrix with large positive diagonal elements. 

\subsection{EM-iterations}
Let us define the collection of the factors over time $F = \{f_t\}_{t=-s}^T$ and the would-be-complete data set $Y = \{\ti{y}_t\}_{t=1}^T$. The issue at hand is to find an estimator of $\theta = \{\Lambda, \Phi, \Psi, \Omega\}$ which maximizes the posterior distribution $p(\theta|Y^A)$, where $Y^A \subseteq Y$ is all available data. This issue is muddled by the fact that $F$ is unobserved and some of $Y$ might be unobserved. A direct optimization can be computationally demanding, especially if the number of variables is large, which is often the case. Instead, we use an EM-algorithm to iteratively find analytically tractable surrogate functions to maximize. 

\cite{DempsterEtAl1977} introduced the EM algorithm for the purpose of ML estimation in case of incomplete data. The principle is to take the expectation of the log-likelihood function with respect to some latent variable, given parameter values from a previous iteration (E-step). Parameters are updated by maximizing the expected log-likelihood (M-step). This procedure is iterated until convergence. 

Due to the fact that the expected likelihood is always equal or smaller than the true likelihood, and exactly equal at its maximum, the procedure converges to a true local maximum or a ridge with a set of local maxima \citep[see][]{Wu1983}. By these properties, EM algorithm is special case of the Minorize-Maximization algorithm, where the expected log-likelihood serve as a surrogate function being a minorized version of the log-likelihood \citep[see][chapter 8.3]{HunterLange2000, OrtegaRheinboldt1970}. In the same way, the expectation of log joint posterior of factors and parameters (with respect to factors) can serve as a surrogate function for the log posterior distribution. Consequently, the procedure is also suitable for MAP estimation \citep[see][p. 26-27]{EMalgorithm2008}.

\par In turn out that we can directly integrate out missing data from the likelihood function in the exact factor model case. Define $Y^M \subseteq Y$ as the missing data, such that $Y^A \cap Y^M = \emptyset$, where $\emptyset$ is the empty set. Further, define $a_{i,t} = \mathbbm{1}\left\{\ti{y}_{i,t} \in Y^A\right\}$, i.e an indicator equal to 1 if $\ti{y}_{i,t}$ is available and 0 otherwise, and $T_i = \sum_{i=1}^T a_{i,t}$ as the number of available observations of variable $i$. Then the likelihood is given by
\begin{align*}
    p\left(Y^A|\theta,F\right) &= \int p\left(Y|\theta,F\right) dY^M = \int \prod_{i=1}^n\prod_{t=1}^T\left(\sqrt{\frac{\psi_i}{2\pi}}\exp\left\{-\frac{\psi_i}{2}\left(\ti{y}_{i,t} - \lambda_i' F_t\right)^2 \right\} \right) dY^M.
\end{align*}
Missing data can be factorized out according to
\begin{align*}
     p\left(Y^A|\theta,F\right) & = \prod_{i=1}^n\prod_{t=1}^T\left[\left(\sqrt{\frac{\psi_i}{2\pi}}\exp\left\{-\frac{\psi_i}{2}\left(\ti{y}_{i,t} - \lambda_i' F_t\right)^2 \right\} \right)^{\mathlarger{a_{i,t}}} \right. \\
    &\qquad \qquad \times \left. \int  \left(\sqrt{\frac{\psi_i}{2\pi}}\exp\left\{-\frac{\psi_i}{2}\left(\ti{y}_{i,t} - \lambda_i' F_t\right)^2 \right\} \right)^{1-\mathlarger{a_{i,t}}} d\ti{y}_{i,t}\right],
\end{align*}
where the integral corresponding to missing data is equal to one, yielding
\begin{align*}
     p\left(Y^A|\theta,F\right) &= \prod_{i=1}^n\left(\left(\frac{\psi_i}{2\pi}\right)^{T_i/2} \exp\left\{-\frac{\psi_i}{2}\sum_{t=1}^T  a_{i,t}\left(\ti{y}_{i,t} - \lambda_i' F_t\right)^2 \right\} \right). \numberthis \label{likelihood}
\end{align*} This means that the only latent variables we have to consider is the unobserved factors.  \cite{BanburaetAl2014} treat missing data as latent variables as well, which is unnecessary in the exact factor model case and makes their algorithm somewhat more convoluted.

We define the joint posterior of parameters and factors as $p\left(F,\theta|Y^A \right)$ and the set of parameters in iteration $k$ as $\theta^{(k)}$. To simply notation we define the expectation operator $\E{(k)}{\cdot} \triangleq \E{F}{\cdot|\theta^{(k)},Y^A}$. The E-step finds the surrogate function
% surrogate function
\begin{align}
    U\left(\theta \big| \theta^{(k)}, Y^A \right) = \E{(k)}{\ln p\left(F, \theta|Y^A\right)}.
\end{align}
This is done practically by running the model by the Kalman filter and smoother, while conditioning on $\theta^{(k)}$. A state space representation of \eqref{ObserveForm2}-\eqref{FactorForm2} is given in Appendix A \footnote{If the number of variables is a a lot larger than the number of states, substantial computational gains can be made by collapsing the vector of observables  according to \cite{JungbackerKoopman2015}.}. The M-step finds parameter values 
% Maximization step 
\begin{align}
    \theta^{(k+1)} = \underset{\theta}{\arg \max} \: U\left(\theta|\theta^{(k)},Y^A\right)
\end{align}
The procedure is iterated until the increase in posterior density between two subsequent step is very small. \footnote{Similar to \cite{BanburaetAl2014} we apply the convergence criteria $\frac{\ln p\left(\theta^{(k)}|Y^A\right)-\ln p\left(\theta^{(k-1)}|Y^A\right)}{1/2\left(\left|\ln p\left(\theta^{(k)}|Y^A\right)\right| + \left|\ln p\left(\theta^{(k-1)}|Y^A\right)\right|\right)} < 10^{-4}$.} By Bayes' theorem we know that
% Log joint posterior 1
\begin{align*}
    \ln p(F,\theta|Y^A) = \ln p(Y^A|F,\theta) + \ln p(F|\theta) + \ln p(\theta) - \ln p(Y^A) \numberthis \label{logjoint1}
\end{align*} Inserting \eqref{likelihood} and prior densities, \eqref{logjoint1} translates into 
% Log joint posterior 2
\begin{align*}
    \ln p(F,\theta|Y^A) &= \sum_{i=1}^n \frac{T_i}{2}\ln \psi_i - \frac{1}{2}\sum_{i=1}^n\sum_{t=1}^T \psi_i a_{i,t} \left(\ti{y}_{i,t} - \lambda_i' F_t^\Lambda \right)^2 \\
    &\quad  - \frac{1}{2}\tilde{F}_0'\Sigma_0 \tilde{F}_0 - \sum_{j=1}^r \frac{T}{2}\ln \omega_j - \frac{1}{2}\sum_{j=1}^r \sum_{t=1}^T \omega_j\left(f_{j,t} - \phi_j' F^{\Phi}_{t-1}\right)^2 \\
    &\quad -\frac{1}{2}\sum_{i=1}^n\lambda_i' V_i^{-1} \lambda_i -\frac{1}{2}\sum_{j=1}^r\phi_j' W_j^{-1}\phi_j - \frac{1}{2}\sum_{i=1}^n\ln \psi_i - \frac{1}{2}\sum_{j=r}^n\ln \omega_j + C, \label{logpost} \numberthis
\end{align*} where $C$ is a constant in terms of $\theta$ and $F$. 

We find  $\theta^{(k+1)}$ by means of derivatives (see Appendix B).  By using the EMC-version of the EM-algorithm \citep{MengRubin1993} the maximization do not need to be applied to all parameters simultaneously; individual parameter blocks can be updated sequentially, fixing the remaining blocks. Starting with $\Phi$, we get M-step
% Partial derivative Phi
\begin{align*}
     \phi_j^{(k+1)} = \left(\sum_{t=1}^T \E{(k)}{F^{\Phi}_{t-1} {F^{\Phi'}_{t-1}}} + \frac{1}{\omega_j^{(k)}} W^{-1}_j\right)^{-1}\Bigg(\E{(k)}{F^{\Phi}_{t-1} f_{j,t}}\Bigg), \quad j = 1, ..., r, \numberthis \label{Phij1}
\end{align*} where we make use of previous iteration $\omega_j^{(k)}$. $\E{(j)}{f_t F_{t-1}^{\Phi'}}$ and $\E{(j)}{F_{t-1}^{\Phi} F_{t-1}^{\Phi'}}$ are obtainable objects from the Kalman smoother in the E-step \citep[see e.g.][]{DeJongMacKinnon1988}. The expression \eqref{Phij1} is similar to traditional multivariate Bayesian linear regression, with the difference that we have expectations due to the latency of the factors. If  there is no shrinkage (i.e $W^{-1}_j = 0$) and $q=1$, this expression is identical to the one given by \cite{BanburaetAl2014}, in matrix form.\footnote{See equation (6) in \cite{BanburaetAl2014}, p.137.} 

\par Similarily, we can find the M-step for $\Lambda$: 
\begin{align*}
     \lambda_i^{(k+1)} = \left(\sum_{t=1}^T a_{i,t} \E{(k)}{F^{\Lambda}_t {F^{\Lambda'}_t}} + \frac{1}{\psi_i^{(k)}} V_i^{-1}\right)^{-1}\left(\sum_{t=1}^T \E{(k)}{F^{\Lambda}_t} a_{i,t} \ti{y}_{i,t}\right), \quad i = 1, ..., n. \numberthis \label{Lambdai1}
\end{align*} \eqref{Lambdai1} builds upon adding and multiplying matrix and vector elements corresponding to available data. If there is no available observations for variable $i$, the only information given is found in the prior, which in this case yields $\lambda^{(k+1)}=0$. Again, if there is no shrinkage (i.e $V^{-1}_i = 0$) and no loading lags ($p=0$), the expression is identical to the one given by \cite{BanburaetAl2014}. \footnote{See expression with vectors stacked in equation (11) in \cite{BanburaetAl2014}, p.138.}

\par Worth noting is that \eqref{Lambdai1} is based on the assumption that $\Psi$ is diagonal. If the assumption does not hold, there may be contemporaneous correlation between available and missing data not contained in the factors, which are disregarded in this expression. 

Turning to  $\Omega$ and $\Psi$, their respective M-steps are given by

\begin{align*}
    \omega_j^{(k+1)} &= \frac{T - 1}{\sum_{t=1}^T \E{(k)}{f_{j,t}^2} - 2 \sum_{t=1}^T\E{(k)}{f_{j,t}F_{t-1}^{\Phi'}}\phi^{(k+1)}_j + \phi^{(k+1)'}_j\left(\sum_{t=1}^T\E{(k)}{ F^{\Phi}_{t-1} F^{\Phi'}_{t-1}}\right)\phi^{(k+1)}_j}, \\
    &\qquad \qquad  j=1, ..., r, \numberthis \label{omegaj} \\ 
    \psi_i^{(k+1)} &= \frac{T_i - 1}{\sum_{t=1}^T a_{i,t} \ti{y}_{i,t}^2 - 2\lambda^{(k+1)'}_i\sum_{t=1}^T\E{(k)}{F_t^\Lambda}a_{i,t}\ti{y}_{i,t} + \lambda^{(k+1)'}_i\left(\sum_{t=1}^Ta_{i,t}\E{(k)}{ F^{\Lambda}_t F^{\Lambda'}_t}\right)\lambda^{(k+1)}_i}, \\
    & \qquad \qquad  i=1, ..., n. \numberthis \label{psii} 
\end{align*}

To satisfy the properties, $0<\omega_j^{(k+1)}$ and $0 < \psi_i^{(k+1)}$, we need $T > 1$ and $T_i > 1$, respectively. Put in other words, we require at least 2 available observations per variable. The denominators in \eqref{omegaj}-\eqref{psii} are the expected sum of square residuals under the assumption that $\Phi$ respectively $\Lambda$ are fixed. In the ML-counterpart, the nominators are instead given by $T$ and $T_i$. Consequently, our proposed MAP estimator does not only shrink $\Lambda$ and $\Phi$, but $\Psi$ and $\Omega$ as well. 
It is worth noting that posterior distributions are not generally invariant to parameter transformation. The choice of parameter functional form, for any parameter, is therefore not an inconsequential decision in MAP estimation. We argue that maximizing in terms of $\Omega$ and $\Psi$ are a reasonable choice, as the resulting estimator is in line with standard expressions for precision estimates. 

\subsection{Adapting factor loading shrinkage} \label{sec.adapt}

Selecting shrinkage hyperparameters $\eta_{\lambda_1}, \eta_{\lambda_2}, ..., \eta_{\lambda_n}$ is not a trivial task. Many approaches for choosing parameter shinkage as been proposeed in BVAR-literature. \cite{Litterman1986} compares different values in an out-of-sample forecasting exercise; \cite{CarrieroEtAl2012} considers shrinkage maximizing the marginal likelihood; \cite{BanburaetAl2010} estimate models over a grid of shrinkage values and choose the value which yields in-sample fit closest to the mean result; and \cite{CarrieroEtAl2015} suggest a full Bayesian treatment with a hierarchical prior structure. All these suggestions require potentially computationally heavy simulation techniques or at least many sequential runs of the model. We suggest keeping the benefit of quick estimation by making the decision part of the EM-algorithm. This is done by introducing hierarchical prior distributions and include shrinkage parameters as latent variables in the E-step. 
The hierarchical prior structure is chosen as
\begin{align*}
    \lambda_i|\eta_{\lambda_i} &\sim \mathcal{N}\left(0, \frac{1}{\eta_{\lambda_i}}J_\Lambda^{-1} \right),  \\
    \eta_{\lambda_i} &\sim \text{Gamma}(\alpha_\Lambda, \beta_\Lambda), \qquad i= 1, ..., n,
\end{align*}
where $\alpha_\Lambda > 0$ and $\beta_\Lambda > 0$. 

\par We consider the EM-algorithm with aggregated set of latent variables $\mathcal{L} = \{F,\eta_{\lambda_1}, ..., \eta_{\lambda_n}\}$. Appendix C points out that the new surrogate function $U^\star\left(\theta|\theta^{(k)},Y^A\right)$ only have linear terms of $\eta_{\lambda_i}$, $\forall i$. Consequently, these hyperparameters can be directly exchanged by their expectations. 

More specifically $V_i$ in \eqref{Lambdai1} can be exchanged in each $(k+1)$th M-step by 
\begin{align*}
    V_i^{(k+1)} = \frac{1}{ \E{(k)}{\eta_{\lambda_i}}}J_\Lambda^{-1}, \quad i = 1, ...,n,
\end{align*}
where we from Appendix C have
\begin{align*}
    \E{(k)}{\eta_{\lambda_i}} = \frac{r(p+1)/2 + \alpha_\Lambda}{{\lambda_i^{(k)}}'J_\Lambda \lambda_i^{(k)}/2 + \beta_\Lambda}, \quad i = 1, ..., n.
\end{align*}

In the limit where $\alpha_\Lambda$ and $\beta_\Lambda$ are zero, the expression reduces to the inverse mean sum of square of lag-decay weighted $\lambda_i^{(k)}$. This represent an adapting scheme, where the hyperparameters take into account the size of the factor loadings, individually for each variable in each M-step, and the chosen lag-decay structure. The procedure is somewhat heuristic and simple, but quick, adapts to a wide range of simulated data sets and rivals ML, as we will show in a Monte Carlo study. 

\section{Monte Carlo study}

This section presents a Monte Carlo excercise to evaluate the estimation capabilities of the MAP estimation algorithm. We assess the estimation precision for a number of different cases, including shares of missing data, sample sizes and number of variables. Similar to \cite{BanburaetAl2014} \citep[see also][]{StockWatson2002a, DozEtAl2011, SolbergerSpanberg2020} we simulate factors $F$ and data $Y$ by 
\begin{align*}
    \ti{y}_t &= \Lambda_0 f_t + \Lambda_1 f_{t-1} + ... + \Lambda_p f_{t-p} + \epsilon_t, \quad \quad \quad \quad &\epsilon_t &\sim \mathcal{N}\left(0, \Sigma\right) \\
    f_t &= \Phi f_{t-1} + \te{u}_t, \; \quad \quad &\te{u}_t &\sim \mathcal{N}\left(0, I_r\right)
\end{align*}

for $t=1, ..., T$, where
\begin{align*}
    \qquad \qquad \qquad \left(\Lambda_{\ell}\right)_{i,z} &\sim N\left(0,1\right), &\Phi_{j,z} &= \begin{cases} \alpha_j \quad \text{for } j = z \\ 0 \quad \text{for } j \neq z\end{cases}\\
    \Sigma_{i,m} &= \delta^{|i-m|}\sqrt{\gamma_i \gamma_m}, \quad &\gamma_i &= \frac{\beta_i}{1-\beta_i} \sum_{\ell=0}^p\sum_{j=1}^r \frac{\left(\Lambda_l\right)_{i,j}^2}{{1 - \alpha_j^2}} \\
    \beta_i &\sim U\left(0.1, 0.9\right), &\alpha_j &\sim U\left(-0.95,0.95\right)
\end{align*}
for $i=1, ..., n$, $j=1, ..., r$, $m=1, ..., n$, $\ell = 0, ..., p$, and $z=1, ..., r$.
Several parameters govern different aspects of the simulation. $\alpha_j$ decides the persistence of factor $j$, $\beta_i$ is the signal-to-noise ratio between idiosyncratic variance and total signal variance for variable $i$ (i.e $\mathbbm{V}[\epsilon_{i,t}]/\mathbbm{V}[\te{y}_{i,t}])$ and $\delta$ denotes idiosyncratic cross-correlation, where $\delta > 0$ violates the diagonal assumption of $\Sigma$ and represents an approximate factor model. In other words, the factor persistence, the signal-to-noise ratio and the particular dynamic of which factors load onto variables are sampled for each factor and variable. 

We simulate factors and data under assumptions $r=1$ or $r=2$ for different number of variables $n$, sample sizes $T$, loading lags $p$ and idiosyncratic cross-correlation $\delta$. 

Thereafter we compare the estimation capabilities of the MAP estimator to ML.\footnote{ML-estimator by \cite{BanburaetAl2014} only assume no loading lags. We will in some cases consider loading lags to make a fair comparison. ML-estimation coincides with the MAP-estimator with 0 shrinkage and $T$ and $T_i$ in variance expression nominator.} We look at different shares of missing data (0\%, 20\% and 40\% respectively) by setting some of the simulated data as missing, chosen by uniform randomization. Additionally, we study what happens when the model is oversaturated with too many factors and/or loading lags. For distinction, $r$ and $p$ are the number of factors and loading lags used in the simulation, whereas $\hat{r}$ and $\hat{p}$ are the counterparts assumed in model estimation. 

The estimation methods are evaluated by their common component estimation errors. We denote the common component $\mathcal{X}_t = \Lambda_0 f_t + \Lambda_1 f_{t-1} + ... + \Lambda_p f_{t-p}$, which is a $n$-length vector $\mathcal{X}_t  = \left[\chi_{1,t} \; ... \; \chi_{n,t}\right]'$. Further, define $s_{y_i}^2$ as the sample variance of variable $i$. Our evaluation statistic is the root mean square error: 
\begin{align}
   \te{RMSE} = \sqrt{\frac{1}{D\times n \times T}\sum_{d=1}^{D}\sum_{i=1}^n \sum_{t=1}^T\frac{\left(\chi_{i,t}^{(d)} - \widehat{\chi}_{i,t}^{(d)}\right)^2}{s^2_{y_i}}},
\end{align}
where $D$ is the number of simulated data sets, $\chi_{i,t}^{(d)}$ is the simulated common component corresponding to $i$th variable element of $\te{y}_{t}$ in data set $d$ and $\widehat{\chi}_{i,t}^{(d)}$ the corresponding model estimates.\footnote{In model estimation, we scale variables to standard deviation 1, to reduce the risk of scaling problems, and then re-scale the common component post-estimation to original scale.} 
We simulate 200 data sets for each parameter set.

\par 
Table \ref{MCr1} shows the resulting RMSE for MAP and the relative RMSE (MAP/ML). For the MAP-estimator we have chosen a small shrinkage on transition parameters $\eta_{\phi_j} = 1/100, \forall j$, and a BVAR-standard lag-decay for factor loadings $(\ell_\Lambda = 2)$. Overall shrinkage for factor loadings are chosen by the adapting scheme according to Section \ref{sec.adapt}. The table is divided into row-blocks according to different sets of parameter values. The first two columns denote number of variables and sample size respectively. Three columns measure RMSE for MAP with $r=1$ over different fractions of missing data, adjacent next three the relative RMSE to ML, which then is reiterated for the last six columns with $r=2$.  When relative RMSE is below 1, MAP is superior to ML, and vice versa. 

We can see a general pattern of growing RMSE with increasing fractions of missing data. Also, RMSE decreases with larger sample sizes and cross-sections. In other words, more data enhances precision. However, idiosyncratic cross-correlation leads to much bigger errors, where the data size and missing patterns does not seem to make any major difference. 

Errors are also bigger when we have more factors, loading lags and/or oversaturated with too many estimated factors. 

In general MAP performs favorably, or at least equal, to ML. This is shown in every case except one $(p=2, n=10, T=100)$, in which MAP have approximately 1\% larger RMSE. We also denote that the relative precision benefits of MAP seems to grow with larger fraction of missing data, insinuating that missing data is one of its comparative advantages. 

The single smallest relative RMSE is shown in an overfitted model with $\hat{r} = 3$ when $r=2$, with high fraction of missing data. This is in line with intuition, as parameter shrinkage is often used to deal with overfitting problems.  

Table 2 show some additional examples where the model are estimated with an overabundance of loading lags. In these examples we also compare to the MAP-estimator without lag decay (i.e $\ell_\Lambda = 0$). The relative RMSE of the latter is given in the last three columns. Lag-decay does not seem to affect the error-sizes generally, although in a few cases, specifically a few specifications with $\hat{r} = r + 1$, it seem to be beneficial. Again, the MAP-estimator provides preferable results to ML, albeit in most cases only marginally. An overabundance of loading lags lead to higher RMSE in general; and more data and less fraction of missing data decrease errors, which is to be expected. 
\pagebreak 

\begin{table}[H]
\renewcommand{\arraystretch}{1.1}
\resizebox{\textwidth}{!}{
\begin{threeparttable}
\caption{Monte Carlo evaluation, common component RMSE} 
\label{MCr1}

\begin{tabular}{rr| rrr r rrr |r rrr r rrr}
\hline \\ [-2ex]											
&	&	\multicolumn{3}{c}{MAP,	$r=1$}	&	&	\multicolumn{3}{c}{$\mathlarger{\frac{\text{MAP}}{\text{ML}}},	r=1$}	&	&	\multicolumn{3}{c}{MAP	$r=2$}	&	&	\multicolumn{3}{c}{$\mathlarger{\frac{\text{MAP}}{\text{ML}}},	r=2$}	\\											
\cmidrule{3-5} \cmidrule{7-9} \cmidrule{11-13} \cmidrule{15-17}	
$n$  & $T$ & $0 \%$ & $20 \%$ & $40 \%$ && $0 \%$ & $20 \%$ & $40 \%$ & & $0 \%$ & $20 \%$ & $40 \%$ && $0 \%$ & $20 \%$ & $40 \%$ \\	
\hline	\\[-2ex]	\multicolumn{17}{c}{$p=0, \quad \delta = 0, \quad  \hat{r} = r, \quad \hat{p} = p,$}	\\	
10	&	50	&	0.25	&	0.28	&	0.32	&&	0.96	&	0.96	&	0.94	&&	0.36	&	0.41	&	0.47	&&	0.97	&	0.96	&	0.95	\\
10	&	100	&	0.19	&	0.21	&	0.24	&&	0.97	&	0.97	&	0.96	&&	0.28	&	0.31	&	0.35	&&	0.98	&	0.97	&	0.96	\\
50	&	50	&	0.20	&	0.22	&	0.27	&&	0.96	&	0.96	&	0.95	&&	0.28	&	0.32	&	0.39	&&	0.98	&	0.98	&	0.96	\\
50	&	100	&	0.15	&	0.17	&	0.20	&&	0.97	&	0.97	&	0.97	&&	0.21	&	0.24	&	0.28	&&	0.98	&	0.97	&	0.97	\\
100	&	50	&	0.19	&	0.21	&	0.25	&&	0.96	&	0.96	&	0.95	&&	0.26	&	0.30	&	0.36	&&	0.98	&	0.97	&	0.96	\\
100	&	100	&	0.14	&	0.16	&	0.19	&&	0.98	&	0.97	&	0.97	&&	0.20	&	0.22	&	0.26	&&	0.97	&	0.97	&	0.96	\\
\\[-2ex]	\multicolumn{17}{c}{$p=0, \quad \delta = 0.5, \quad  \hat{r} = r, \quad \hat{p} = p,$}	\\		
10	&	50	&	0.66	&	0.66	&	0.65	&&	0.98	&	0.98	&	0.97	&&	0.70	&	0.71	&	0.71	&&	0.99	&	0.99	&	0.98	\\
10	&	100	&	0.67	&	0.66	&	0.65	&&	0.99	&	0.99	&	0.99	&&	0.69	&	0.69	&	0.68	&&	0.99	&	1.00	&	0.99	\\
50	&	50	&	0.69	&	0.69	&	0.69	&&	0.99	&	0.98	&	0.98	&&	0.72	&	0.72	&	0.73	&&	1.00	&	1.00	&	1.00	\\
50	&	100	&	0.69	&	0.69	&	0.69	&&	0.99	&	0.99	&	0.99	&&	0.71	&	0.71	&	0.71	&&	1.00	&	1.00	&	1.00	\\
100	&	50	&	0.69	&	0.69	&	0.69	&&	0.99	&	0.98	&	0.98	&&	0.72	&	0.73	&	0.74	&&	1.00	&	1.00	&	1.00	\\
100	&	100	&	0.69	&	0.69	&	0.69	&&	0.99	&	0.99	&	0.99	&&	0.71	&	0.71	&	0.71	&&	1.00	&	1.00	&	1.00	\\
\\[-2ex]	\multicolumn{17}{c}{$p=0, \quad \delta = 0, \quad  \hat{r} = r+1, \quad \hat{p} = p,$}	\\																									
10	&	50	&	0.37	&	0.41	&	0.47	&&	0.98	&	0.96	&	0.95	&&	0.45	&	0.51	&	0.60	&&	0.96	&	0.96	&	0.95	\\
10	&	100	&	0.29	&	0.31	&	0.35	&&	0.99	&	0.97	&	0.97	&&	0.36	&	0.39	&	0.44	&&	0.98	&	0.97	&	0.96	\\
50	&	50	&	0.28	&	0.32	&	0.38	&&	0.98	&	0.97	&	0.96	&&	0.34	&	0.40	&	0.49	&&	0.99	&	0.99	&	0.98	\\
50	&	100	&	0.21	&	0.24	&	0.28	&&	0.98	&	0.97	&	0.97	&&	0.26	&	0.30	&	0.36	&&	0.99	&	0.98	&	0.99	\\
100	&	50	&	0.27	&	0.30	&	0.36	&&	0.98	&	0.97	&	0.97	&&	0.32	&	0.37	&	0.46	&&	0.99	&	0.99	&	0.99	\\
100	&	100	&	0.20	&	0.22	&	0.26	&&	0.97	&	0.97	&	0.96	&&	0.24	&	0.28	&	0.33	&&	0.98	&	0.99	&	0.99	\\
\\[-2ex]	\multicolumn{17}{c}{$p=2, \quad \delta = 0, \quad  \hat{r} = r, \quad \hat{p} = p,$}	\\																									
10	&	50	&	0.39	&	0.44	&	0.53	&&	0.99	&	0.97	&	0.95	&&	0.56	&	0.66	&	0.85	&&	0.98	&	0.97	&	0.96	\\
10	&	100	&	0.30	&	0.34	&	0.39	&&	1.01	&	1.00	&	0.98	&&	0.44	&	0.50	&	0.62	&&	1.01	&	0.99	&	0.98	\\
50	&	50	&	0.32	&	0.37	&	0.45	&&	0.99	&	0.99	&	0.98	&&	0.45	&	0.54	&	0.67	&&	0.99	&	0.99	&	0.98	\\
50	&	100	&	0.24	&	0.28	&	0.33	&&	0.99	&	0.99	&	0.99	&&	0.34	&	0.40	&	0.48	&&	0.99	&	0.99	&	0.99	\\
100	&	50	&	0.31	&	0.35	&	0.42	&&	0.99	&	0.99	&	0.99	&&	0.43	&	0.50	&	0.63	&&	0.99	&	0.99	&	0.99	\\
100	&	100	&	0.23	&	0.26	&	0.31	&&	0.99	&	0.99	&	0.99	&&	0.32	&	0.37	&	0.45	&&	0.99	&	0.99	&	0.99	\\
\\[-2ex]	\multicolumn{17}{c}{$p=2, \quad \delta = 0.5, \quad  \hat{r} = r, \quad \hat{p} = p,$}	\\																									
10	&	50	&	0.69	&	0.70	&	0.71	&&	0.99	&	0.99	&	0.98	&&	0.77	&	0.81	&	0.90	&&	0.99	&	0.99	&	0.98	\\
10	&	100	&	0.68	&	0.68	&	0.68	&&	1.00	&	1.00	&	0.99	&&	0.73	&	0.74	&	0.77	&&	1.00	&	0.99	&	0.99	\\
50	&	50	&	0.72	&	0.72	&	0.73	&&	1.00	&	1.00	&	0.99	&&	0.75	&	0.78	&	0.82	&&	1.00	&	1.00	&	0.99	\\
50	&	100	&	0.71	&	0.71	&	0.71	&&	1.00	&	1.00	&	1.00	&&	0.73	&	0.74	&	0.76	&&	1.00	&	1.00	&	1.00	\\
100	&	50	&	0.71	&	0.72	&	0.73	&&	1.00	&	1.00	&	0.99	&&	0.75	&	0.77	&	0.81	&&	1.00	&	1.00	&	1.00	\\
100	&	100	&	0.71	&	0.71	&	0.71	&&	1.00	&	1.00	&	1.00	&&	0.73	&	0.74	&	0.76	&&	1.00	&	1.00	&	1.00	\\
\\[-2ex]	\multicolumn{17}{c}{$p=2, \quad \delta = 0, \quad  \hat{r} = r+1, \quad \hat{p} = p,$}	\\																									
10	&	50	&	0.57	&	0.67	&	0.84	&&	0.98	&	0.97	&	0.95	&&	0.70	&	0.87	&	1.06	&&	0.99	&	0.98	&	0.95	\\
10	&	100	&	0.44	&	0.50	&	0.62	&&	1.00	&	0.99	&	0.98	&&	0.56	&	0.65	&	0.85	&&	1.00	&	0.99	&	1.00	\\
50	&	50	&	0.45	&	0.54	&	0.68	&&	0.99	&	0.99	&	0.99	&&	0.55	&	0.67	&	0.82	&&	0.99	&	0.99	&	0.93	\\
50	&	100	&	0.34	&	0.40	&	0.48	&&	0.99	&	0.99	&	0.99	&&	0.42	&	0.49	&	0.62	&&	0.99	&	1.00	&	1.00	\\
100	&	50	&	0.43	&	0.50	&	0.63	&&	0.99	&	0.99	&	0.99	&&	0.52	&	0.63	&	0.79	&&	0.99	&	0.99	&	0.97	\\
100	&	100	&	0.32	&	0.37	&	0.45	&&	0.99	&	0.99	&	0.99	&&	0.40	&	0.46	&	0.56	&&	0.99	&	0.99	&	1.00	\\
\\[-2ex]	\hline													
\end{tabular}
    \begin{tablenotes}
      \small
      \item \textit{Remarks:} MAP denotes the Maximum-a-posteriori estimation models with lag-decay prior and ML denotes maximum likelihood estimation. $0 \%$, $20 \%$ and $40 \%$ denotes the respective shares of missing data. $n$ is cross-section size and $T$ is sample size. $r$ and $p$ are the number of factors and factor lags in true model, respectively, with corresponding estimation counterparts $\hat{r}$ and $\hat{p}$. $\delta$ affects idiosyncratic cross-correlation.
    \end{tablenotes}
\end{threeparttable}}
\end{table}

\begin{table}[H]
\centering
\begin{threeparttable}
\caption{Monte Carlo evaluation, common component RMSE} 
\label{MCr2}
\begin{tabular}{rr| rrr r rrr r rrr}
\hline \\ [-2ex]								
&	&	\multicolumn{3}{c}{MAP,	}	&	&	\multicolumn{3}{c}{$\mathlarger{\frac{\text{MAP}}{\text{ML}}},	$}	&	&	\multicolumn{3}{c}{$\mathlarger{\frac{\text{MAP}}{\text{MAP no lag-decay}}}	$}	\\	
\cmidrule{3-5} \cmidrule{7-9} \cmidrule{11-13}	
$n$  & $T$ & $0 \%$ & $20 \%$ & $40 \%$ && $0 \%$ & $20 \%$ & $40 \%$ & & $0 \%$ & $20 \%$ & $40 \%$ \\	
\hline	\\[-2ex]	\multicolumn{13}{c}{$r=1, \quad p=0,  \quad  \hat{r} = r, \quad \hat{p} = p+3$}	\\
10	&	50	&	0.47	&	0.55	&	0.70	&&	0.98	&	0.97	&	0.97	&&	1.00	&	0.99	&	1.00	\\
10	&	100	&	0.35	&	0.41	&	0.50	&&	0.99	&	0.99	&	0.98	&&	1.00	&	1.00	&	1.00	\\
50	&	50	&	0.40	&	0.46	&	0.57	&&	0.99	&	0.98	&	0.98	&&	1.00	&	1.00	&	1.00	\\
50	&	100	&	0.30	&	0.34	&	0.41	&&	0.99	&	0.99	&	0.99	&&	1.00	&	1.00	&	1.00	\\
100	&	50	&	0.38	&	0.44	&	0.54	&&	0.99	&	0.99	&	0.99	&&	1.00	&	1.00	&	1.00	\\
100	&	100	&	0.28	&	0.32	&	0.39	&&	0.99	&	0.99	&	0.99	&&	1.00	&	1.00	&	1.00	\\
\\[-2ex]	\multicolumn{13}{c}{$r=2, \quad p=0,  \quad  \hat{r} = r, \quad \hat{p} = p+3$}	\\	
10	&	50	&	0.62	&	0.76	&	0.93	&&	0.98	&	0.98	&	0.93	&&	1.00	&	0.99	&	0.96	\\
10	&	100	&	0.48	&	0.56	&	0.72	&&	0.99	&	0.99	&	0.99	&&	1.00	&	1.00	&	1.00	\\
50	&	50	&	0.51	&	0.61	&	0.76	&&	0.99	&	0.99	&	0.96	&&	1.00	&	1.00	&	0.98	\\
50	&	100	&	0.39	&	0.45	&	0.55	&&	0.99	&	0.99	&	0.99	&&	1.00	&	1.00	&	1.00	\\
100	&	50	&	0.48	&	0.57	&	0.72	&&	0.99	&	0.99	&	0.98	&&	1.00	&	1.00	&	1.00	\\
100	&	100	&	0.37	&	0.42	&	0.51	&&	0.99	&	0.99	&	1.00	&&	1.00	&	1.00	&	1.00	\\
\\[-2ex]	\multicolumn{13}{c}{$r=1, \quad p=2,  \quad  \hat{r} = r, \quad \hat{p} = p+2$}	\\	
10	&	50	&	0.46	&	0.55	&	0.69	&&	0.98	&	0.97	&	0.97	&&	1.00	&	0.99	&	0.99	\\
10	&	100	&	0.35	&	0.41	&	0.50	&&	1.00	&	0.99	&	0.98	&&	1.00	&	1.00	&	1.00	\\
50	&	50	&	0.40	&	0.46	&	0.57	&&	0.99	&	0.99	&	0.98	&&	1.00	&	1.00	&	1.00	\\
50	&	100	&	0.30	&	0.34	&	0.41	&&	0.99	&	0.99	&	0.99	&&	1.00	&	1.00	&	1.00	\\
100	&	50	&	0.38	&	0.44	&	0.54	&&	0.99	&	0.99	&	0.99	&&	1.00	&	1.00	&	1.00	\\
100	&	100	&	0.28	&	0.32	&	0.39	&&	0.99	&	0.99	&	0.99	&&	1.00	&	1.00	&	1.00	\\
\\[-2ex]	\multicolumn{13}{c}{$r=1, \quad p=2,  \quad  \hat{r} = r+1, \quad \hat{p} = p+2$}	\\	
10	&	50	&	0.66	&	0.79	&	0.97	&&	0.98	&	0.94	&	0.90	&&	0.99	&	0.96	&	0.94	\\
10	&	100	&	0.52	&	0.62	&	0.81	&&	0.99	&	0.99	&	0.99	&&	1.00	&	1.00	&	1.00	\\
50	&	50	&	0.55	&	0.67	&	0.77	&&	0.99	&	0.99	&	0.86	&&	1.00	&	1.00	&	0.91	\\
50	&	100	&	0.42	&	0.49	&	0.62	&&	0.99	&	0.99	&	1.00	&&	1.00	&	1.00	&	1.00	\\
100	&	50	&	0.53	&	0.64	&	0.79	&&	0.99	&	0.99	&	0.95	&&	1.00	&	1.00	&	0.97	\\
100	&	100	&	0.40	&	0.46	&	0.57	&&	0.99	&	0.99	&	1.00	&&	1.00	&	1.00	&	1.00	\\
\\[-2ex]	\hline	
\end{tabular}
    \begin{tablenotes}
      \small
      \item \textit{Remarks:} MAP denotes the Maximum-a-posteriori estimation models with lag-decay prior and ML denotes maximum likelihood estimation. $0 \%$, $20 \%$ and $40 \%$ denotes the respective shares of missing data. $n$ is cross-section size and $T$ is sample size. $r$ and $p$ are the number of factors and factor lags in true model, respectively, with corresponding estimation counterparts $\hat{r}$ and $\hat{p}$.
    \end{tablenotes}
\end{threeparttable}
\end{table}

\section{Conclusions}
This paper introduces a method of maximum a posteriori estimation for dynamic factor models with incomplete data, and evaluates its estimation precision in a Monte Carlo study. We show how to modify and expand upon the EM-iterations of \cite{BanburaetAl2014} by including parameter shrinkage in a Minnesota style prior fashion. Additionally, we present a heuristic, quick and simple method of factor loading shrinkage selection, imbedded in the EM-algorithm. We evaluate the estimator with and without lags of factors in state space signal equations by measuring RMSE of common components, and compare the results to the ML estimator. 

The study suggests that MAP estimation is preferable to ML in general and increasingly so with larger fractions of missing data. The results seem to hold over different sample sizes, number of variables, loading lags and number of factors. They also suggest that MAP-estimation is preferable in case of overfitting, when models are estimated with more factors and/or lags than given by the true process. 

Our presented method can provide quick factor estimates and predictions in practical forecasting settings. It can be applied in small and large scale models, in case of mixed frequencies, different variable sample sizes and publication delays. The method might also be a practical tool in missing data imputation or backtracking missing time series. We further believe it can be extended in several different ways. For example, there are several other parameter shrinkage techniques to be considered, which is left for future research. We suggest researchers to continue to investigate fast and practical algorithms for large scale factor models, who also apply parameter shrinkage. 

\bibliography{references}
\newpage

\section*{Appendix A: State space formulation}
This section shows the state space formulation of \eqref{ObserveForm2}-\eqref{Fphi}. The state vector is given by $\tilde{F}_t = \begin{bmatrix} f_t' & f_{t-1}' & ... & f_{t-s}'\end{bmatrix}'$, where $s = \max \left(p,q-1\right)$. In which case the model can be written as:

% State space formulation
\begin{align*}
\ti{y}_t &= \left[
    \begin{array}{c;{2pt/2pt}c}
    \underset{n \times r(p+1)}{\Lambda} & \underset{n \times r(s-p)}{0}
    \end{array} \right]\tilde{F}_t + \epsilon_t, \quad \epsilon_t \sim \mathcal{N}\left(0,\Psi\right) \\ \tilde{F}_t &= \left[
    \begin{array}{cccc}
    & & & \multicolumn{1}{;{2pt/2pt}c}{}  \\[-2ex]
      \multicolumn{3}{c}{\underset{r \times rq}{\Phi}} & \multicolumn{1}{;{2pt/2pt}c}{\underset{r \times r(s-q+1)}{0}} \\ \hdashline[2pt/2pt]
       \multicolumn{2}{c}{\underset{rs \times rs}{I}} & \multicolumn{2}{;{2pt/2pt}c}{\underset{rs \times r}{0}}
    \end{array}
\right]\tilde{F}_{t-1} + \left[
    \begin{array}{c}
    \underset{r \times r}{I} \\ \hdashline[2pt/2pt]
    \underset{rs \times r}{0}
    \end{array}
\right]\xi_t, \quad \xi_t \sim \mathcal{N}\left(0, \Omega\right)
\end{align*}

If the length of $\ti{y}_t$ is much larger than $s$, then substantial computational gains can be made by collapsing $\ti{y}_t$ by the technique of \cite{JungbackerKoopman2015}.  

\section*{Appendix B: EM derivations} \label{sec.B}
This section derives the EM-iteration expressions \eqref{Phij1}-\eqref{psii}, which show similarities to derivations of \cite{BanburaetAl2014}.

To find the expression for $\Phi^{(k+1)}$ we take the partial derivative of $U\left(\theta \big| \theta^{(k)}, Y^A\right)$ with respect to $\phi_j$, which is given by
\begin{align*}
    &\frac{\partial}{\partial \phi_j} U\left(\theta|\theta^{(k)}, Y^A \right) = \frac{\partial}{\partial \phi_j}\left(-\frac{1}{2} \sum_{t=1}^T \omega_j \E{(k)}{\left(f_{j,t} - \phi_j' F^{\Phi}_{t-1}\right)^2} - \frac{1}{2} \phi_j' W^{-1}_j \phi_j \right) \\ 
    &\qquad =  -\frac{1}{2}\frac{\partial}{\partial \phi_j}\Bigg( -2\sum_{t=1}^T \omega_j \phi_j' \E{(k)}{F^{\Phi}_{t-1} f_{j,t}}  + \sum_{t=1}^T \omega_j \phi_j'\E{(k)}{F^{\Phi}_{t-1} {F^{\Phi'}_{t-1}}} \phi_j   + \phi_j' W^{-1}_j \phi_j \Bigg) \\
    &\qquad =  \omega_j\sum_{t=1}^T \E{(k)}{F^{\Phi}_t f_{j,t}} - \omega_j\sum_{t=1}^T  \E{(k)}{F^{\Phi}_{t-1} {F^{\Phi'}_{t-1}}} \phi_j -  W^{-1}_j \phi_j \\
   &\qquad =  \omega_j\left(\sum_{t=1}^T \E{(k)}{F^{\Phi}_t f_{j,t}} - \left(\sum_{t=1}^T  \E{(k)}{F^{\Phi}_{t-1} {F^{\Phi'}_{t-1}}} +  \omega_j^{-1}W^{-1}_j\right) \phi_j \right).
\end{align*}
Setting the derivative to zero and rearranging terms:
\begin{align*}
     \phi_j^{(k+1)} = \left(\sum_{t=1}^T \E{(k)}{F^{\Phi}_{t-1} {F^{\Phi'}_{t-1}}} + \omega_j^{-1} W^{-1}_j\right)^{-1}\Bigg(\E{(k)}{F^{\Phi}_{t-1} f_{j,t}}\Bigg).
\end{align*}

Turning to $\Lambda$:
\begin{align*}
    &\frac{\partial}{\partial \lambda_i} U\left(\theta|\theta^{(k)}, Y^A \right) = \frac{\partial}{\partial \lambda_i}\left(-\frac{1}{2} \sum_{t=1}^T \psi_i a_{i,t} \E{(k)}{\left(\ti{y}_{i,t} - \lambda_i' F^{\Lambda}_t\right)^2} - \frac{1}{2} \lambda_i' V_i^{-1} \lambda_i \right) \\ 
    &\qquad =  -\frac{1}{2}\frac{\partial}{\partial \lambda_i}\Bigg( -2\sum_{t=1}^T \psi_i \lambda_i' \E{(k)}{F^{\Lambda}_t} a_{i,t} \ti{y}_{i,t} + \sum_{t=1}^T \psi_i a_{i,t} \lambda_i'\E{(k)}{F^{\Lambda}_t {F^{\Lambda'}_t}} \lambda_i   + \lambda_i'V_i^{-1} \lambda_i \Bigg) \\
    &\qquad =  \psi_i \sum_{t=1}^T \E{(k)}{F^{\Lambda}_t} a_{i,t}  \ti{y}_{i,t} - \psi_i \sum_{t=1}^T a_{i,t} \E{(k)}{F^{\Lambda}_t {F^{\Lambda'}_t}} \lambda_i -  V_i^{-1} \lambda_i \\
   &\qquad = \psi_i\left(\sum_{t=1}^T \E{(k)}{F^{\Lambda}_t} a_{i,t}  \ti{y}_{i,t} - \left(\sum_{t=1}^T a_{i,t} \E{(k)}{F^{\Lambda}_t {F^{\Lambda'}_t}} +  \psi_i^{-1}V_i^{-1}\right) \lambda_i \right).
\end{align*}
Setting the derivative to zero and rearranging terms:
\begin{align*}
     \lambda_i^{(k+1)} = \left(\sum_{t=1}^T a_{i,t} \E{(k)}{F^{\Lambda}_t {F^{\Lambda'}_t}} + \psi_i^{-1} V_i^{-1}\right)^{-1}\left(\sum_{t=1}^T \E{(k)}{F^{\Lambda}_t} a_{i,t} \ti{y}_{i,t}\right).
\end{align*}
Turning to $\Omega$:
\begin{align*}
    &\frac{\partial}{\partial \omega_j} U\left(\theta|\theta^{(k)}, Y^A \right) = \frac{\partial}{\partial \omega_j}\left(\frac{T}{2}\ln \omega_j - \frac{1}{2} \sum_{t=1}^T \omega_j \E{(k)}{\left(f_{j,t}  - \phi_j' F^{\Phi}_{t-1}\right)^2} - \frac{1}{2} \ln \omega_j \right) \\
    &\qquad = \frac{1}{2}\frac{\partial}{\partial \omega_j}\left((T - 1)\ln \omega_j - \omega_j \left(\sum_{t=1}^T \E{(k)}{f_{j,t}^2} - 2 \sum_{t=1}^T\E{(k)}{f_{j,t}F_{t-1}^{\Phi'}}\phi_j \right. \right. \\
    &\qquad \qquad \qquad \qquad \qquad \qquad \qquad \qquad \left. \left. + \phi_j'\left(\sum_{t=1}^T\E{(k)}{ F^{\Phi}_{t-1} F^{\Phi'}_{t-1}}\right)\phi_j \right)  \right) \\
    &\qquad = \frac{1}{2}\left(\frac{T - 1}{\omega_j} - \left(\sum_{t=1}^T \E{(k)}{f_{j,t}^2} - 2 \sum_{t=1}^T\E{(k)}{f_{j,t}F_{t-1}^{\Phi'}}\phi_j \right. \right. \\
    &\qquad \qquad \qquad \qquad \qquad \qquad \qquad \qquad \left. \left. + \phi_j'\left(\sum_{t=1}^T\E{(k)}{ F^{\Phi}_{t-1} F^{\Phi'}_{t-1}}\right)\phi_j \right)  \right).
\end{align*}

Setting derivative to zero yields\begin{align*}
    \omega_j^{(k+1)} = \frac{T - 1}{\sum_{t=1}^T \E{(k)}{f_{j,t}^2} - 2 \sum_{t=1}^T\E{(k)}{f_{j,t}F_{t-1}^{\Phi'}}\phi_j + \phi_j'\left(\sum_{t=1}^T\E{(k)}{ F^{\Phi}_{t-1} F^{\Phi'}_{t-1}}\right)\phi_i}.
\end{align*}
Lastly, turning to $\Psi$:
\begin{align*}
    \frac{\partial}{\partial \psi_i} U\left(\theta|\theta^{(k)}, Y^A \right) &= \frac{\partial}{\partial \psi_i}\left(\frac{T_i}{2}\ln \psi_i - \frac{1}{2} \sum_{t=1}^T \psi_i a_{i,t} \E{(k)}{\left(\ti{y}_{i,t} - \lambda_i' F^{\Lambda}_t\right)^2} - \frac{1}{2} \ln \psi_i \right) \\
    &= \frac{1}{2}\frac{\partial}{\partial \psi_i}\left((T_i - 1)\ln \psi_i - \psi_i \left(\sum_{t=1}^T a_{i,t} \ti{y}_{i,t}^2 - 2 \lambda_i'\sum_{t=1}^T\E{(k)}{F_t^\Lambda}a_{i,t}\ti{y}_{i,t} \right. \right. \\
    &\qquad \qquad \qquad \qquad \qquad \qquad \qquad \qquad \left. \left. + \lambda'\left(\sum_{t=1}^Ta_{i,t}\E{(k)}{ F^{\Lambda}_t F^{\Lambda'}_t}\right)\lambda_i \right)  \right) \\
    &= \frac{1}{2}\left(\frac{T_i - 1}{\psi_i} - \left(\sum_{t=1}^T a_{i,t} \ti{y}_{i,t}^2 - 2 \lambda_i'\sum_{t=1}^T\E{(k)}{F_t^\Lambda}a_{i,t}y_{i,t} \right. \right. \\
    &\qquad \qquad \qquad \qquad \qquad \qquad \qquad \qquad \left. \left. + \lambda_i'\left(\sum_{t=1}^Ta_{i,t}\E{(k)}{ F^{\Lambda}_t F^{\Lambda'}_t}\right)\lambda_i \right)  \right).
\end{align*}

Setting derivative to zero yields
\begin{align*}
    \psi_i^{(j+1)} = \frac{T_i - 1}{\sum_{t=1}^T a_{i,t} \ti{y}_{i,t}^2 - 2\lambda_i'\sum_{t=1}^T\E{(k)}{F_t^\Lambda}a_{i,t}\ti{y}_{i,t} + \lambda_i'\left(\sum_{t=1}^Ta_{i,t}\E{(k)}{ F^{\Lambda}_t F^{\Lambda'}_t}\right)\lambda_i}
\end{align*}
\newpage
\section*{Appendix C: Hyperparameter iterations}
We first note that as $\eta_{\lambda_1}$, ..., $\eta_{\lambda_n}$ and $F$ share no common terms in $\ln p(\mathcal{L}, \theta|Y^A)$, as $\eta_{\lambda_i}$ only appears in the parameter prior part $p(\lambda_i,\eta_{\lambda_i})$ individually for each $i$. Conclusively, we can deduce independence according to $p(\mathcal{L}|Y^A,\theta) = p(F|Y^A,\theta)\prod_{i=1}^n p(\eta_{\lambda_i}|Y^A,\theta)$.  As such, the addition of $p(\eta_{\lambda_i})$ does not change the functional form of $p(F|Y^A,\theta)$, nor do we have to take the expectations in the E-step over $F$ and $\eta_{\lambda_i}$ jointly. The only term in $U^\star\left(\theta|\theta^{(k)},Y^A\right) = \E{\mathcal{L}|\theta^{(k)}}{\ln p(\mathcal{L},\theta|Y_A)}$ that $\theta$ and $\eta_{\lambda_i}$ share is $-\eta_{\lambda_i}\frac{1}{2}\sum_{i=1}^n \lambda_i'J_\Lambda \lambda_i$, which is linear in terms of $\eta_{\lambda_i}$. Thus, $\eta_{\lambda_i}$ can be exchanged in the maximization step by $\E{\eta_{\lambda_i}|\theta^{(k)}}{\eta_{\lambda_i}}$. We can find the expectation by noting that
\begin{align*}
p(\eta_{\lambda_i}|Y^A,\theta) &\underset{\eta_{\lambda_i}}{\propto} p(F,\theta,\eta_{\lambda_1}, ...,\eta_{\lambda_n}|Y^A) \underset{\eta_{\lambda_i}}{\propto} \det\left(\eta_{\lambda_i} J_\Lambda\right)^{1/2} \exp\left\{-\frac{1}{2} \lambda_i' \eta_{\lambda_i} J_\Lambda \lambda_i \right\} \eta_{\lambda_i}^{\alpha_\Lambda - 1} \exp\left\{- \beta_\Lambda \eta_{\lambda_i} \right\} \\
&\underset{\eta_{\lambda_i}}{\propto} \eta_{\lambda_i}^{\frac{r(p+1)}{2} + \alpha_\Lambda - 1}\exp\left\{-\left(\frac{\lambda_i'J_\Lambda \lambda_i}{2} + \beta_\Lambda\right)\eta_{\lambda_i}\right\} \\
&\underset{\eta_{\lambda_i}}{\propto} \text{Gamma}\left(\frac{r(p+1)}{2} + \alpha_\Lambda, \frac{\lambda_i'J_\Lambda \lambda_i}{2} + \beta_\Lambda\right).
\end{align*}

Consequently, by Gamma distribution property, we have
\begin{align*}
    \E{\eta_{\lambda_i}|\theta^{(k)}}{\eta_{\lambda_i}} = \frac{r(p+1)/2 + \alpha_\Lambda}{{\lambda_i^{(k)}}'J_\Lambda \lambda_i^{(k)}/2 + \beta_\Lambda}.  
\end{align*}

\end{document}